# The temperature dependent thermal potential in Quantum Boltzmann equation

Zheng-Chuan Wang

The University of Chinese Academy of Sciences, P. O. Box 4588, Beijing 100049, China, wangzc@ucas.ac.cn

## Abstract

To explore the thermal transport procedure driven by temperature gradient in terms of linear response theory, Luttinger et al. proposed the thermal scalar and vector potential[1,2]. In this manuscript, we try to address the microscopic origin of these phenomenological thermal potentials. Based on the temperature dependent damping force derived from quantum Boltzmann equation (QBE), we express the thermal scalar and vector potential by the distribution function in damping force, which originates from the scattering of conduction electrons. We illustrate this by the scattering of electron-phonon interaction in some systems. The temperature and temperature gradient in the local equilibrium distribution function will have effect on the thermal scalar and vector potentials, which is compatible with the previous works[1,2]. The influence from quantum correction terms in QBE are also considered, which contribute not only to the damping force, but also to the anomalous velocity in the drift term. An approximated solution for the QBE is given, the numerical results for the damping force, thermal current density as well as other physical observable are shown in figures.

## I. Introduction

Classical Boltzmann equation was widely used to explore the transport procedure not only in dilute gas[3], but also in conduction electrons of normal metal, or in quasi-particles as well as elementary excitation of strongly interacting systems[4]. However, it fails to describe the quantum transport in microscopic systems, because the position $\vec{r}$ and momentum $\vec{p}$ in it's distribution function can not be determined simultaneously due to the uncertainty principle, so Wigner made a quantum transformation on the lesser Green function $G^<$, and introduced a new distribution function to overcome this trouble[5], which satisfies the QBE. QBE had also been derived by Kadanoff and Baym based on non-equilibrium Green function formalism[6], Mahan and Hansch further gave out the QBE including electric and magnetic field[7]. It should be pointed out that the Wigner distribution function can not merely be interpreted as the probability distribution function like classical distribution function, it has negative value due to the quantum correlation therein. To surmount the difficulty of negative probability, Husimi proposed another distribution function by performing coarse average over the Wigner distribution and obtain a positive distribution function finally[8], but the position accuracy had been lowered after this coarse average. In 2008, Wang et al. expanded the Wigner distribution function by the Planck constant $\hbar$[9], where the zero order term corresponds to the positive classical distribution function, while the first and higher order terms can be treated as the quantum corrections on the classical distribution function, which may have negative value, both the zero order distribution function and higher order quantum distribution functions have the same position accuracy as the Wigner distribution function, so our method can keep the position accuracy of zero and higher order distribution functions in addition to obtain the positive zero order distribution function. Under this expansion, the quantum entropy can naturally reduce to the classical entropy when we let $\hbar \rightarrow 0$ in the classical limit.

Besides driven by the electric field or magnetic field, the transport procedure can also be driven by the temperature gradient. Whatever the temperature or the thermal force which is proportional to the temperature gradient are all macroscopic quantities, one can not include the temperature gradient into a microscopic Hamiltonian like the mechanical case, and the thermal conductivity can not be directly calculated as the electrical conductivity by the linear response of the system to an external electric field, too. In order to conveniently explore the thermal transport coefficients by Kubo formula mechanically, in 1964 Luttinger introduced a scalar function $\Psi$ to describe the effect of temperature gradient in the Hamiltonian of a thermal transport system[1], then the thermal transport coefficient could be obtained by the linear response theory with respect to the field $\Psi$. Despite of the widely applications of scalar $\Psi$ on various thermally induced electronic transport[10-16], the non-physical divergence were found for the transport coefficient when T → $0$[10,11,14], then Tatara proposed a thermal vector potential to replace Luttinger's scalar potential in the Hamiltonian to avoid this divergence[2], in which the equilibrium diamagnetic current were treated consistently. However, whatever the thermal scalar or vector potential, their microscopic origin had not been addressed clearly, in this manuscript, we will explore this issue.

In 2023, Wang derived a temperature dependent damping force from the classical Boltzmann equation[17], where the thermal damping force originates from the Compton scattering of transport particles with the radiation photons of reservoir. This thermal damping force can also be derived from the scattering of transport particles and the reservoir particles with mass in the framework of Vlasov equation, the inverse damping relaxation time and non-equilibrium temperature were defined, too[18]. It is shown that the thermal damping force can be related to the thermal scalar and vector potential, then we can find the microscopic origin of these thermal potential following the above works. In this manuscript, we will investigate the temperature dependent damping force in the QBE, then express the thermal damping force by the thermal scalar and vector potential, and explore the microscopic expression of the thermal potential.

**II. Theoretical formalism**

Consider the electronic transport under an external electric field $U(\vec{r})$ in a conductor, the QBE satisfied by the conduction electrons can be written as

$$\frac{\partial f(\vec{p},\vec{r},\omega,t)}{\partial t} + \vec{v}\cdot\vec{\nabla}f(\vec{p},\vec{r},\omega,t) + \frac{\partial U(\vec{r})}{\partial \vec{r}}\cdot$$

$$\vec{\nabla}_p f(\vec{p},\vec{r},\omega,t) = -2\pi[\Sigma^> G^< - \Sigma^< G^>]-$$

$$2\pi i\hbar[\Sigma^<, ReG^r] - 2\pi i\hbar[ReG^r, \Sigma^<], \quad (1)$$

where $f(\vec{p},\vec{r},\omega,t)$ is the quantum Wigner distribution function, $\Sigma^>$ and $\Sigma^<$ are the greater and lesser self-energies, $G^>$ and $G^<$ are the greater and lesser Green functions, respectively, the bracket $[A,B]$, i.e. $[\Sigma^<, ReG^r]$, is defined as $\frac{\partial A}{\partial \vec{r}}\frac{\partial B}{\partial \vec{p}} - \frac{\partial A}{\partial \vec{p}}\frac{\partial B}{\partial \vec{r}}$, $G^< = if(\vec{p},\vec{r},\omega,t)$ and $G^> = if'(\vec{p},\vec{r},\omega,t)$, where $f'(\vec{p},\vec{r},\omega,t)$ is the distribution function of hole. If we only consider the electron-phonon interaction in the system, the self-energy can be written as[7]

$$\Sigma^<(\vec{p},\omega) = \int\frac{d^3q}{(2\pi)^3}M_q^2\int\frac{d\omega}{2\pi}[(N_q+1)G^<(\vec{p}+\vec{q},\omega+\omega_q)+N_q G^<(\vec{p}+\vec{q},\omega-\omega_q)], \quad (2)$$

and

$$\Sigma^>(\vec{p},\omega) = \int\frac{d^3q}{(2\pi)^3}M_q^2\int\frac{d\omega}{2\pi}[N_q G^>(\vec{p}+\vec{q},\omega+\omega_q)+(N_q+1)G^>(\vec{p}+\vec{q},\omega-\omega_q)], \quad (3)$$

where, $N_q$ is the phonon occupation number, $M_q$ stands for the matrix element of electron-phonon interaction, $\omega_q$ is the frequency of phonon with momentum $\vec{q}$, then the first term in the right hand of Eq.(1) can be expressed as

$$-2\pi\{(i\int\frac{d^3\vec{q}}{(2\pi)^3}M_q^2\int\frac{d\omega}{2\pi}af(\vec{p}+$$

$\vec{q},\omega))f(\vec{p},\vec{r},\omega,t) - (i\int\frac{d^3\vec{q}}{(2\pi)^3}M_q^2\int\frac{d\omega}{2\pi}bf(\vec{p}+\vec{q},\omega))f'(\vec{p},\vec{r},\omega,t))\}$, (4)

where $a = \frac{(N_q+1)}{E-\omega-\omega_q+i\delta} + \frac{N_q}{E-\omega+\omega_q+i\delta}$, $b = \frac{N_q}{E-\omega-\omega_q+i\delta} + \frac{(N_q+1)}{E-\omega+\omega_q+i\delta}$. If the momentum transfer $\vec{q}$ caused by the electron-phonon scattering is small compared with the momentum $\vec{p}$ of electron, we can make a Taylor series expansion on the distribution function $f'(\vec{p}+\vec{q},\omega)$ and $f(\vec{p}+\vec{q},\omega)$ around $\vec{p}$ in the above integral, then Eq.(4) can be written as

$$-2\pi i[f'f\int\frac{d^3\vec{q}}{(2\pi)^3}M_q^2\int\frac{d\omega}{2\pi}(\frac{1}{E-\omega-\omega_q+i\delta} - \frac{1}{E-\omega+\omega_q+i\delta})]$$
$$-2\pi i[(\int\frac{d^3\vec{q}}{(2\pi)^3}M_q^2\int\frac{d\omega}{2\pi}\vec{q}a)\frac{\partial f'}{\partial \vec{p}}f - (\int\frac{d^3\vec{q}}{(2\pi)^3}M_q^2\int\frac{d\omega}{2\pi}\vec{q}b)\cdot\frac{\partial f}{\partial \vec{p}}f'),\quad(5)$$

If we neglect the second and third terms which are the small quantum corrections in the right hand of Eq.(1), we can rewrite Eq.(1) as

$$\frac{\partial f(\vec{p},\vec{r},\omega,t)}{\partial t} + \vec{v}\cdot\vec{\nabla}f(\vec{p},\vec{r},\omega,t) + [\frac{\partial U(\vec{r})}{\partial \vec{r}} + 2\pi i(\int\frac{d^3\vec{q}}{(2\pi)^3}M_q^2\int\frac{d\omega}{2\pi}\vec{q}b)f']\cdot\vec{\nabla}_p f(\vec{p},\vec{r},\omega,t) = -2\pi i[(\int\frac{d^3\vec{q}}{(2\pi)^3}M_q^2\int\frac{d\omega}{2\pi}\vec{q}a)\cdot\frac{\partial f'}{\partial \vec{p}}f,\quad(6)$$

Since $N_q = e^{-\beta\hbar\omega_q}$, where $\beta=1/k_BT$, the term $2\pi i(\int\frac{d^3\vec{q}}{(2\pi)^3}M_q^2\int\frac{d\omega}{2\pi}\vec{q}b)f'$ is temperature dependent which originates from the electron-phonon interaction, we name it as the damping force, $2\pi i[(\int\frac{d^3\vec{q}}{(2\pi)^3}M_q^2\int\frac{d\omega}{2\pi}a)\vec{q}\cdot\frac{\partial f'}{\partial p}$ refers to the inverse damping relaxation time as given in Ref.[17]. The temperature dependent damping force is a thermal force, so it may be expressed by the thermal scalar and vector potential as

$$\frac{\partial \vec{A}}{\partial t} + \nabla\varphi = 2\pi i(\int\frac{d^3\vec{q}}{(2\pi)^3}M_q^2\int\frac{d\omega}{2\pi}\vec{q}b)f',\quad(7)$$

We can see that the thermal vector potential $\vec{A}$ and the scalar potential $\varphi$ depend on the temperature, they are determined by the Wigner distribution function. When we make a gauge transformation $\varphi \to \varphi - \dot{\chi}$ and $\vec{A} \to \vec{A} + \nabla\chi$, where $\chi$ is a scalar function, Eq.(7) can keep the gauge invariant.

In the steady state, the Wigner distribution function $f'(\vec{p},\vec{r},\omega,t)$ doesn't depend on time, then $\frac{\partial \vec{A}}{\partial t}=0$, and

$$\nabla\varphi = 2\pi i(\int\frac{d^3\vec{q}}{(2\pi)^3}M_q^2\int\frac{d\omega}{2\pi}\vec{q}b)f',\quad(8)$$

so we can obtain the scalar potential as $\varphi = \int 2\pi i(\int\frac{d^3\vec{q}}{(2\pi)^3}M_q^2\int\frac{d\omega}{2\pi}b)f'\vec{q}\cdot d\vec{r}$, while in the special case that the Wigner distribution function doesn't depend on position, we have $\nabla\varphi = 0$, then

$$\frac{\partial \vec{A}}{\partial t} = 2\pi i(\int\frac{d^3\vec{q}}{(2\pi)^3}M_q^2\int\frac{d\omega}{2\pi}\vec{q}b)f',\quad(9)$$

integrating it over time, we can get the thermal vector potential as

$$\vec{A}(t) = \int 2\pi i(\int\frac{d^3\vec{q}}{(2\pi)^3}M_q^2\int\frac{d\omega}{2\pi}\vec{q}b)f'dt,\quad(10)$$

Generally, the Wigner distribution function depends on both the position and time, in order to obtain the thermal scalar and vector potential, we can expand the $f'(\vec{p},\vec{r},\omega,t)$ around local equilibrium distribution function $f_0(\vec{p},\vec{r})$ as

$$f'(\vec{p},\vec{r},\omega,t) = f_0(\vec{p},\vec{r}) + f_1'(\vec{p},\vec{r},\omega,t) + \ldots, \tag{11}$$

where $f_1'(\vec{p},\vec{r},\omega,t)$ is the first order term deviating from the local equilibrium distribution function $f_0(\vec{p},\vec{r})$ which doesn't depend on time $t$. Substituting Eq.(11) into Eq.(7), we have

$$\frac{\partial \vec{A}}{\partial t} + \nabla \varphi =$$
$$2\pi i (\int \frac{d^3\vec{q}}{(2\pi)^3} M_q^2 \int \frac{d\omega}{2\pi} \vec{q} b) f_0(\vec{p},\vec{r}) +$$
$$2\pi i (\int \frac{d^3\vec{q}}{(2\pi)^3} M_q^2 \int \frac{d\omega}{2\pi} \vec{q} b) f_1'(\vec{p},\vec{r},\omega,t). \tag{12}$$

If we simply choose

$$\nabla \varphi = 2\pi i (\int \frac{d^3\vec{q}}{(2\pi)^3} M_q^2 \int \frac{d\omega}{2\pi} \vec{q} b) f_0(\vec{p},\vec{r}), \tag{13}$$

and

$$\frac{\partial \vec{A}}{\partial t} = 2\pi i (\int \frac{d^3\vec{q}}{(2\pi)^3} M_q^2 \int \frac{d\omega}{2\pi} \vec{q} b) f_1'(\vec{p},\vec{r},\omega,t), \tag{14}$$

We can obtain a special solution $\varphi = \int 2\pi i (\int \frac{d^3\vec{q}}{(2\pi)^3} M_q^2 \int \frac{d\omega}{2\pi} b) f_0(\vec{p},\vec{r}) \vec{q} \cdot d\vec{r}$ and

$$\vec{A}(t) = \int 2\pi i (\int \frac{d^3\vec{q}}{(2\pi)^3} M_q^2 \int \frac{d\omega}{2\pi} \vec{q} b) f_1'(\vec{p},\vec{r},\omega,t) dt$$

from Eqs.(13) and (14), which can be regarded as a special gauge chosen for the thermal scalar and vector potential. Since the temperature is position dependent in the local equilibrium distribution function, if we write $T(x) = T_0 + bx$, where b is the temperature gradient, $T_0$ is the temperature at the boundary of the system, then the thermal scalar and vector potential are concerning with the temperature gradient according to Eq.(12), which is consistent with the thermal potential proposed by Luttinger and Tatara[1,2].

In the next, we will further investigate the contributions from quantum correction terms. Similar to the zero order term in the above, the quantum correction terms in the right hand of Eq.(1) can also be made a Taylor series expansion, to the first order it can be written as

$$[-2\pi\hbar i \int \frac{d^3\vec{q}}{(2\pi)^3} M_q^2 \int \frac{d\omega}{2\pi} \vec{q} \cdot \frac{\partial ReG^r}{\partial \vec{p}}$$
$$+2\pi\hbar i (\int \frac{d^3\vec{q}}{(2\pi)^3} M_q^2 \int \frac{d\omega}{2\pi} a) \frac{\partial G^>}{\partial \vec{p}}] \cdot \frac{\partial f}{\partial \vec{r}}$$
$$-[2\pi\hbar i (\int \frac{d^3\vec{q}}{(2\pi)^3} M_q^2 \int \frac{d\omega}{2\pi} b) \frac{\partial ReG^r}{\partial \vec{r}} +$$
$$2\pi\hbar i (\int \frac{d^3\vec{q}}{(2\pi)^3} M_q^2 \int \frac{d\omega}{2\pi} a) \frac{\partial G^>}{\partial \vec{r}}] \cdot \frac{\partial f}{\partial \vec{p}}. \tag{15}$$

Then the QBE which contain the quantum correction terms can be written as:

$$\frac{\partial f(\vec{p},\vec{r},\omega,t)}{\partial t} + (\vec{v}+\vec{C}) \cdot \vec{\nabla} f(\vec{p},\vec{r},\omega,t) +$$
$$[\frac{\partial U(\vec{r})}{\partial \vec{r}} + \vec{D} + \vec{A}f'] \cdot \vec{\nabla}_p f(\vec{p},\vec{r},t) = \vec{B} \frac{\partial f'}{\partial p} f + E, \tag{16}$$

where $\vec{A} = 2\pi i (\int \frac{d^3\vec{q}}{(2\pi)^3} M_q^2 \int \frac{d\omega}{2\pi} \vec{q} b)$, $\vec{B} = -2\pi i [(\int \frac{d^3\vec{q}}{(2\pi)^3} M_q^2 \int \frac{d\omega}{2\pi} \vec{q} a)$,
$\vec{C} = -2\pi\hbar i (\int \frac{d^3\vec{q}}{(2\pi)^3} M_q^2 \int \frac{d\omega}{2\pi} b) \frac{\partial ReG^r}{\partial \vec{p}} +$
$2\pi\hbar i (\int \frac{d^3\vec{q}}{(2\pi)^3} M_q^2 \int \frac{d\omega}{2\pi} a) \frac{\partial G^>}{\partial \vec{p}}$, $\vec{D} = -2\pi\hbar i (\int \frac{d^3\vec{q}}{(2\pi)^3} (M_q^2 \int \frac{d\omega}{2\pi} b) \frac{\partial ReG^r}{\partial \vec{r}} -$
$2\pi\hbar i (\int \frac{d^3\vec{q}}{(2\pi)^3} M_q^2 \int \frac{d\omega}{2\pi} a) \frac{\partial G^>}{\partial \vec{r}})$, $E = -2\pi i [f' f \int \frac{d^3\vec{q}}{(2\pi)^3} M_q^2 \int \frac{d\omega}{2\pi} (\frac{1}{E-\omega-\omega_q+i\delta} - \frac{1}{E-\omega+\omega_q+i\delta})]$. So the quantum correction terms can also contribute to the damping force, moreover, it will contribute to the velocity, too, which is similar to the anomalous velocity in anomalous Hall effect[19].

### III. The solution of temperature dependent QBE

It is difficult to obtain an analytical solution for Eq.(6), below we try to solve it by the method of separating variables. As an example, we only consider the one-dimensional transport simply. Since the electrons transport through conduction band, usually the band width is very narrow in the one-dimensional system, the energy of conduction electrons can be regarded approximately as unchanged, so we don't consider the variable ω in the distribution function, and write the Wigner distribution function simply as:

$$f(p,x,t) = P(p)X(x)T(t), \quad (17)$$

Substituting it into Eq.(6), we have

$$\frac{\partial X(x)}{\partial x}\frac{1}{X(x)} = \frac{[-P(p)\frac{\partial T(t)}{\partial t} - (\frac{\partial U(x)}{\partial x} + Af_0(p,x)T(t))\frac{\partial P(p)}{\partial p} + B\frac{\partial f_0}{\partial p}P(p)T(t)]}{\frac{p}{m}P(p)T(t)}, \quad (18)$$

If we approximate $P(p)$ in the right hand of Eq.(18) as the local equilibrium distribution $f_0$, and $T(t)$ as $e^{-\lambda t}$, where $\lambda$ is the inverse of relaxation time, then we have

$$X(x) = exp[\int dx \frac{[-P(p)\frac{\partial T(t)}{\partial t} - (\frac{\partial U(x)}{\partial x} + Af_0(p,x)T(t))\frac{\partial P(p)}{\partial p} + B\frac{\partial f_0}{\partial p}P(p)T(t)]}{\frac{p}{m}P(p)T(t)}], \quad (19)$$

Similarly, we can write

$$\frac{\partial P(p)}{\partial p}\frac{1}{P(p)} = \frac{[BX(x)T(t)\frac{\partial f_0}{\partial p} - X(x)\frac{\partial T(t)}{\partial t} - \frac{p}{m}T(t)\frac{\partial X(x)}{\partial x}]}{(\frac{\partial U(x)}{\partial x} + Af_0(p,x))X(x)T(t)}, \quad (20)$$

If we substitute $X(x)$ as Eq.(19) and $T(t) = e^{-\lambda t}$ into the above equation, we have

$$P(p) = exp[\int dp \frac{[BX(x)T(t)\frac{\partial f_0}{\partial p} - X(x)\frac{\partial T(t)}{\partial t} - \frac{p}{m}T(t)\frac{\partial X(x)}{\partial x}]}{(\frac{\partial U(x)}{\partial x} + Af_0(p,x))X(x)T(t)}], \quad (21)$$

Also, the time dependent function $T(t)$ can be expressed as

$$\frac{\partial T(t)}{\partial t}\frac{1}{T(t)} = \frac{[B\frac{\partial f_0}{\partial p}P(p)X(x) - \frac{p}{m}P(p)\frac{\partial X(x)}{\partial x} - (\frac{\partial U(x)}{\partial x} + Af_0(p,x))X(x)\frac{\partial P(p)}{\partial p}]}{P(p)X(x)}, \quad (22)$$

If we substitute the expression $X(x)$ in Eq.(19) and $P(p)$ in Eq.(21) into the above equation, we can obtain $T(t)$ as

$$T(t) = exp[\int dt \frac{[B\frac{\partial f_0}{\partial p}P(p)X(x) - \frac{p}{m}P(p)\frac{\partial X(x)}{\partial x} + (\frac{\partial U(x)}{\partial x} + Af_0(p,x))X(x)\frac{\partial P(p)}{\partial p}]}{P(p)X(x)}], \quad (23)$$

Finally, we approximately obtain the Wigner distribution function according to Eq.(17), where $X(x)$, $P(p)$ and $T(t)$ are expressed by Eq.(19), Eq.(21) and (23), respectively.

The physical observable, i.e., the charge density and charge current et al., can be evaluated by the above approximate solution, we will demonstrate it numerically. For simplicity, we only consider the electronic transport through a one-dimensional conductor under an external bias, in which the electron-phonon interaction cause the scattering of conduction electrons.

There exists temperature gradient $k = 2K/nm$ in this system, the position dependent temperature is adopted as: $T(x) = T_0 + kx$, where $T_0$ is temperature at the boundary. In our calculation, the electric field is chosen as $E = -1.0 \times 10^7 V \cdot m^{-1}$. In Fig.1, we plot the damping force as a function of position which is temperature dependent, the higher of the temperature, the bigger of the damping force. The magnitude of damping force is comparable with the external electric field force, so we can not neglect it. The damping force originates from the scattering terms in QBE, in our system it comes from the scattering of electron-phonon, which is a part of electron-phonon interaction, but it is temperature dependent. The charge current density vs position is shown in Fig.2, it increases with position, because the electric field is applied along the -x axis, the current density will decreases along the -x axis due to the electron-phonon scattering, which is the origin of electric resistance in this system. In Fig.3, we draw the charge density vs position, it increases with position which is similar to the charge current density, the maximum of charge density corresponds to the maximum of current density. The thermal current density as a function of position is shown in Fig.4 at different temperatures, the solid line corresponds to the thermal current density at 200K, the dashed line corresponds to the temperature at 250K, while the dotted line corresponds to 300K, the thermal current density increases with temperature, in our study the thermal current is carried by the electrons, so it is analogous to the electronic current.

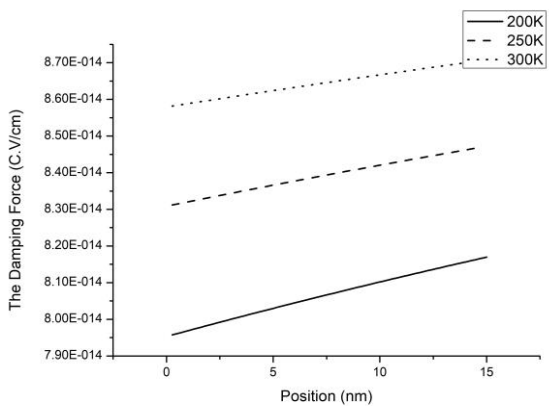

Fig.1 The damping force vs position at temperatures $T_0 = 200K, 250K, 300K$.

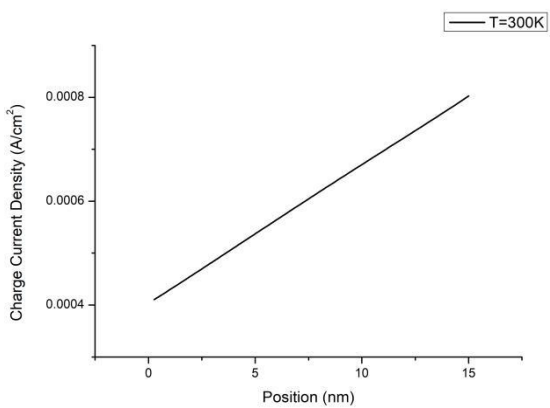

Fig.2 The charge current density vs position at temperature $T_0 = 300K$.

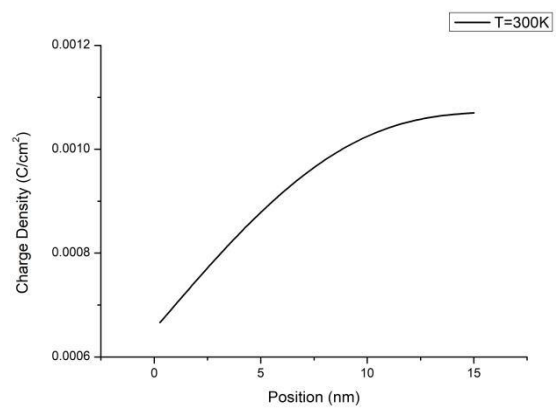

Fig.3 The charge density vs position at temperature $T_0 = 300K$.

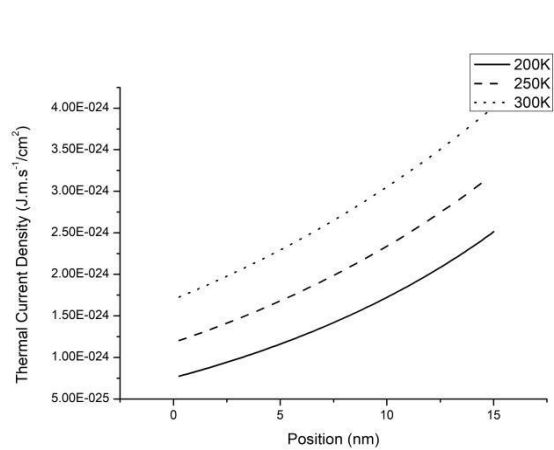

Fig.4 The thermal current density vs position at temperatures $T_0$ =200K, 250K, 300K.

### IV. Summary and discussions

Based on the temperature dependent damping force derived from QBE, we give out a microscopic expression for the thermal scalar and vector potential by use of distribution function, which originates from the scattering of electron-phonon interaction. Certainly, if we include other interactions, i.e., the electron-electron interaction or the electron-impurity interaction, in the self-energy, they have contribution on the thermal damping force, or the thermal scalar and vector potential, too. So the thermal potentials come from the scattering terms in the right hand of QBE. In our manuscript, the damping force, the charge current, the charge density as well as the thermal current density are demonstrated in Fig.1-Fig.4, respectively, they are all temperature dependent. On the other hand, if we include the spin freedom in the QBE, we will obtain the extended spinor Boltzmann equation, which is a powerful tool to study the spin-polarized transport in spintronics, how to express the thermal scalar and vector potential in spinor Boltzmann equation is still an open question, we leave it for future exploration.

### Acknowledgments

This study is supported by the National Key R&D Program of China (Grant No. 2022YFA1402703).

### Data Availability Statement

Data sets generated during the current study are available from the corresponding author on reasonable request.

### Additional information

Competing interest statement: The authors declare that they have no competing interests.